# The science case for a far-infrared interferometer in the era of JWST and ALMA

David Leisawitz*[a], Matteo Bonato[b], Duncan Farrah[c], T. Tupper Hyde[a], Aláine Lee[c], Joshua Bennett Lovell[d], Brenda Matthews[e], Lee G. Mundy[f], Conor Nixon[a], Petr Pokorny[a,g], Berke V. Ricketti[h], Giorgio Savini[i], Jeremy Scott[h], Irene Shivaei[j], Locke Spencer[h], Kate Su[k], C. Megan Urry[l], David Wilner[d]

[a]Goddard Space Flight Center, 8800 Greenbelt Rd., Greenbelt, MD USA 20771
[b]National Institute of Astrophysics (INAF)-Istituto di Radioastronomia, Bologna, Italy
[c]University of Hawai'i at Mānoa, Honolulu County, Hawaii, USA 96822
[d]Harvard-Smithsonian Center for Astrophysics, 60 Garden St., Cambridge, MA USA 02138
[e]National Research Council of Canada, Herzberg Astronomy & Astrophysics Research, Victoria, BC, Canada
[f]University of Maryland, College Park, MD USA 20742
[g]The Catholic University of America, Washington, DC USA 20064
[h]University of Lethbridge, Lethbridge, Alberta T1K 3M4, Canada
[i]University College London, Watford Way 553, Mill Hill, London NW7 2QS, UK
[j]Centro de Astrobiología (CAB), CSIC-INTA, Ctra. de Ajalvir km 4, Torrejón de Ardoz, E-28850, Madrid, Spain
[k]The University of Arizona and Steward Observatory, Tucson, AZ USA 85721
[l]Yale University, P.O. Box 208120, New Haven, CT USA 06520-8120

## ABSTRACT

A space-based far-infrared interferometer could work synergistically with the James Webb Space Telescope (JWST) and the Atacama Large Millimeter Array (ALMA) to revolutionize our understanding of the astrophysical processes leading to the formation of habitable planets and the co-evolution of galaxies and their central supermassive black holes. Key to these advances are measurements of water in its frozen and gaseous states, observations of astronomical objects in the spectral range where most of their light is emitted, and access to critical diagnostic spectral lines, all of which point to the need for a far-infrared observatory in space. The objects of interest – circumstellar disks and distant galaxies – typically appear in the sky at sub-arcsecond scales, which rendered all but a few of them unresolvable with the successful and now-defunct 3.5-m *Herschel Space Observatory*, the largest far-infrared telescope flown to date. A far-infrared interferometer with maximum baseline length in the tens of meters would match the angular resolution of JWST at 10x longer wavelengths and observe water ice and water-vapor emission, which ALMA can barely do through the Earth's atmosphere. Such a facility was conceived and studied two decades ago. Here we revisit the science case for a space-based far-infrared interferometer in the era of JWST and ALMA and summarize the measurement capabilities that will enable the interferometer to achieve a set of compelling scientific objectives. Common to all the science themes we consider is a need for sub-arcsecond image resolution.

# 1. INTRODUCTION

JWST and ALMA give astrophysicists superlative sensitivity and angular resolution with which they probe the universe in search of answers to profound and enduring questions, one of which, put simply, is How did we get here? How did a nearly structureless universe initially devoid of heavy elements develop a rich variety of galaxies and provide life-enabling conditions on a small, rocky planet orbiting an average star in 13.8 billion years?

Key puzzle pieces in the cosmic origin story will remain beyond our grasp until a far-infrared (far-IR; ~25-400 μm) observatory with measurement capabilities comparable to those of JWST and ALMA is available (Figure 1). Here we present a synopsis of the science case for a space-based far-IR interferometer based on three papers[1-3] submitted mid-2023 to the Astrophysical Journal. In section 2, we discuss the need for a far-IR interferometer to understand how planets form and some become habitable, by resolving the structures and mass budgets of essential molecular ingredients for life, such as water ice, water vapor, Hydrogen-, Oxygen-, Carbon-, Sulfur-, and Nitrogen-bearing species. Section 3 discusses mature planetary systems like our own, and how structures in their orbiting dust clouds can reveal currently unseen planets the way gaps in Saturn's rings reveal the presence of moons. Section 4 discusses the niche for a far-IR interferometer in discovering how galaxies evolved. In section 5, we summarize measurement requirements for a far-IR interferometer and compare the predicted performance of a well-studied interferometer, the Space Infrared Interferometric Telescope (SPIRIT)[4], with JWST, ALMA, and past far-IR missions.

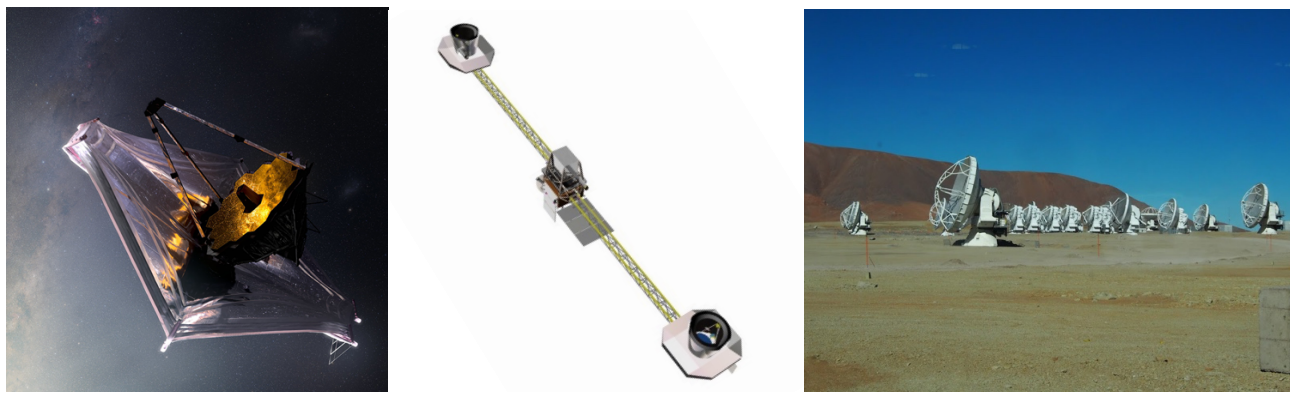

Figure 1. The James Webb Space Telescope (JWST; left), a far-IR interferometer concept based on the SPIRIT design (middle), and the Atacama Large Millimeter Array (ALMA; right). JWST and ALMA are operational and provide unprecedented sensitivity and angular resolution in the near-to-mid-infrared and submillimeter-to-millimeter wavelength ranges, respectively. A far-IR interferometer would offer similar measurement capabilities in the information-rich 25-400 μm wavelength range.

# 2. PLANET FORMATION AND HABITABILITY

ALMA and past far-IR telescopes have enabled a basic understanding of the planet formation process and the structural characteristics of circumstellar protoplanetary disks (Figure 2), and JWST is contributing new insights, yet gaping holes in our knowledge remain. Presently we have no reliable measure of the total masses of protoplanetary disks and the mass distribution in the disks, which serves as the raw material for planet formation. Most of the mass is warm or cold molecular hydrogen ($H_2$), which cannot be observed directly. The indirect tracers of $H_2$ currently available to us, such as CO spectral lines observable with ALMA, are fraught with calibration uncertainties. For example, we do not know how much of the CO is frozen out onto dust grain surfaces and non-emissive, nor the precise ratio of CO to $H_2$. Deuterated hydrogen (HD) is a very good tracer of its lighter cousin, $H_2$, and its low-level rotational lines appear in the far-IR at 56 and 112 μm[5-7].

The water distribution in protoplanetary disks can make the difference between habitable and lifeless planets, and this distribution is a second great unknown. Much of the water is shielded from stellar radiation by dust or lies at too great a distance from the star to exist in gaseous form. Water ice produces broad spectral features in the 40 to 70-μm range (Figure 3) and its spatial distribution can be measured with a far-IR interferometer (Figure 4).

The spectrum of a protoplanetary disk (Figure 3) is also an excellent source of information about evolving chemistry. "It is particularly interesting to understand the budget of C, N, O, and S incorporated into planet-forming gas and solids, as the inventory of these volatile elements will control the atmospheric composition and potential habitability of newly formed planets."[2] These elements can appear in planet atmospheres and affect the planet's potential habitability. Some of the brightest spectral lines that indicate the presence of C, N, O, and S appear in the far-IR, complementing submillimeter observations with ALMA (Figure 3).

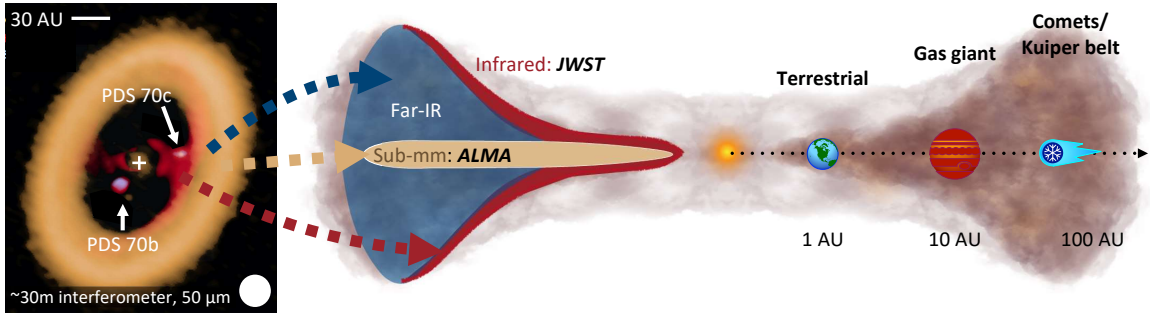

Figure 2. Illustration showing the structure of a protoplanetary disk and the regions probed by ALMA (disk midplane), JWST (disk surface) and in the far-infrared (most of the disk volume). A far-IR interferometer would resolve protoplanetary disks to map their mass distribution, which affects planet development, and the water reservoir available to make some planets habitable. Credits: ALMA (ESO/NRAO/NAOJ) & A. Isella/ESO; M.K. McClure.

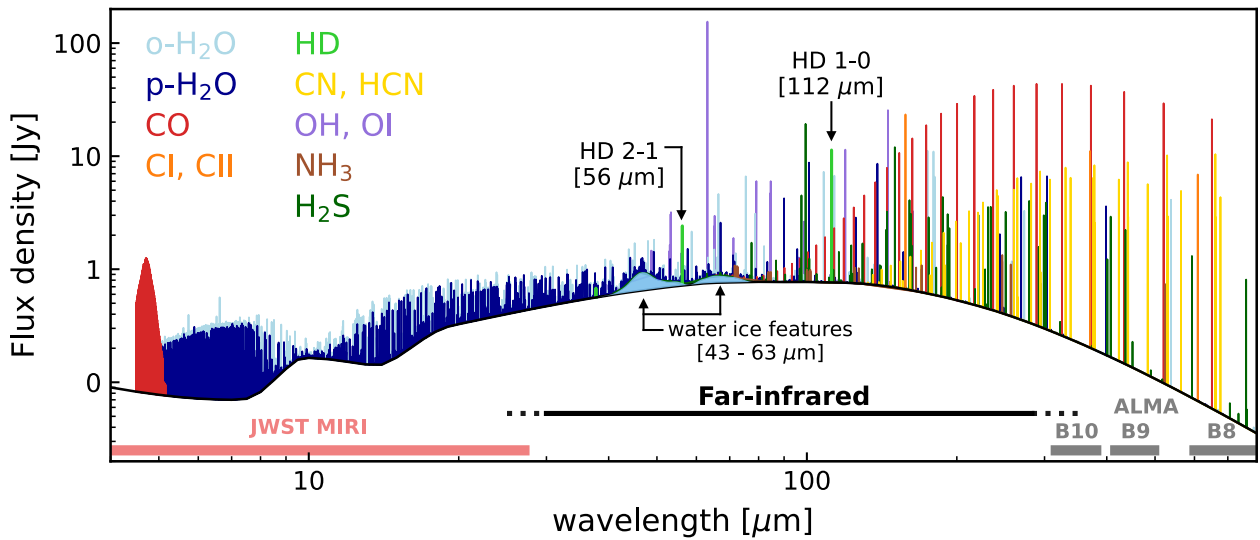

Figure 3. Model spectrum of a protoplanetary disk from [2] showing the brightest "CHNOS" spectral lines in the wavelength range 4-800 μm. Two water ice features are seen in the 40-70 μm range and many water vapor ($H_2O$) spectral lines appear throughout the far-IR. The deuterated hydrogen (HD) lines at 56 and 112 μm are thought to be the best available tracers of protoplanetary disk mass. Nearly all the $H_2O$ line emission is absorbed by water molecules in our atmosphere, putting the submillimeter water lines out of ALMA's reach. Spectroscopic observations with JWST and ALMA provide complementary information, such as JWST measurements of ices in the dense interstellar medium surrounding young protostars.[8]

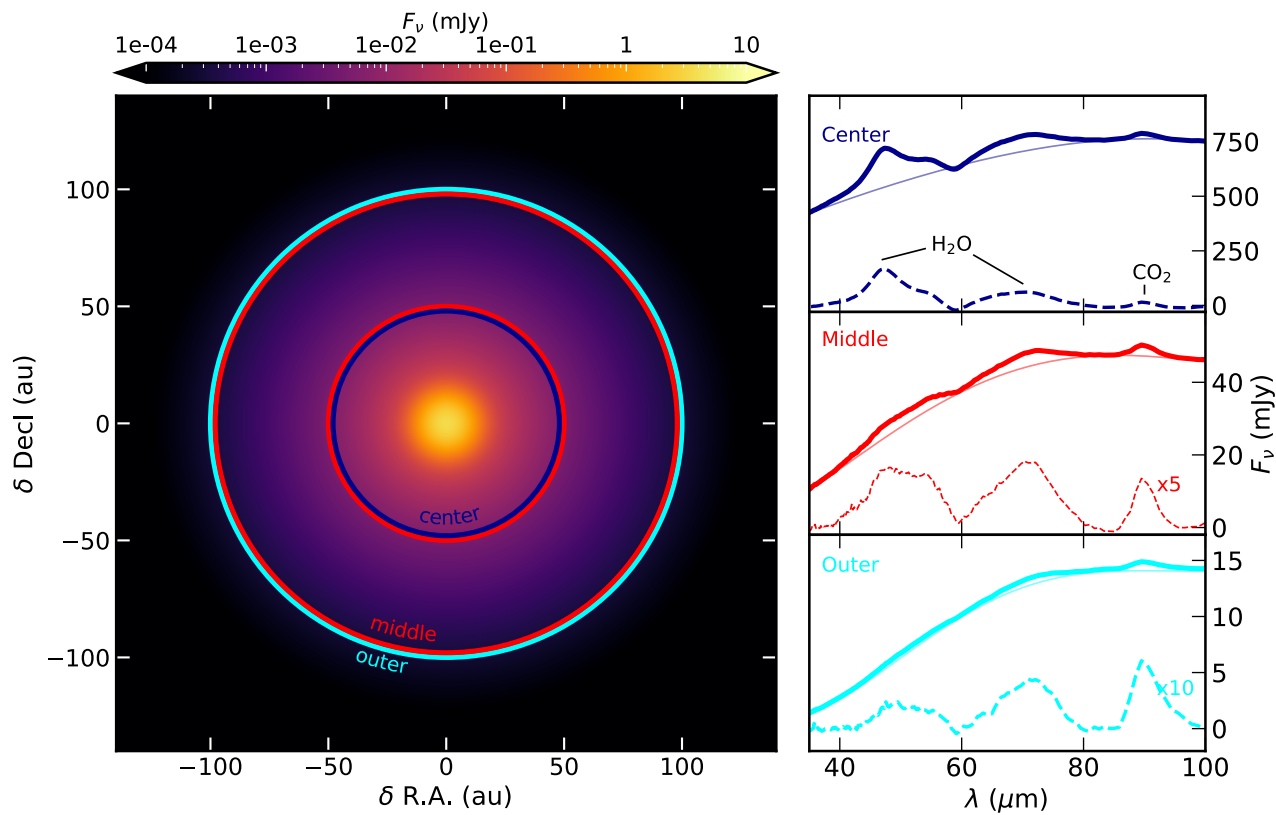

Figure 4. Model from [2] of a representative protoplanetary disk shown face-on at a wavelength of 40 μm (left) and predicted spectra of the center, middle, and outer regions of the disk. A far-IR interferometer could measure the expected radial gradient in the $H_2O$ ice feature strength to map out the water reservoir. The thin solid line shown in each spectrum represents the thermal dust continuum, and the continuum-subtracted spectrum is shown with dashed lines (scaled in the graphs for the middle and outer regions).

## 3. PLANETARY SYSTEM ARCHITECTURES

Mature planetary systems like the Solar System have used up or expelled their natal gas and dust, but occasional collisions between small bodies release "debris," which orbits along with the planets, forming a debris disk. This constantly replenished supply of interplanetary dust is warmed by starlight and glows in the thermal infrared; interplanetary dust in the Solar System is responsible for "zodiacal emission." The dust-releasing planetesimals interact gravitationally with planets, forming resonant structures.[9] The asteroid and Kuiper belts in the Solar System are fossil records of past migrations of Jupiter and Neptune, respectively.

We can reverse-engineer the structures seen in extrasolar debris disks to infer the presence and properties of planets, including Neptune-mass planets at large orbital distances from their host stars (>5 au), which current planet search techniques can hardly touch. Indeed, multiple techniques are needed to gain a complete picture of planetary system architectures because each method comes with its own observational biases. How common is the arrangement of planets in the Solar System? From a complete sample we will learn about planet formation and the evolution of planetary systems, including planet migration.

Very few debris disks are close enough for their structure to have been seen with past infrared telescopes; sub-arcsecond resolution is essential to resolve disks out to 100 pc (Figure 5), thereby expanding the sample size to >100 and enabling us to probe different formation pathways through trend analysis. For example, a large enough sample can be binned by age to look for signs of systematic evolution. ALMA has sufficient resolution, but only a few exceptionally bright debris disks are detectable because ALMA observes them at wavelengths where they are orders of magnitude fainter than their far-IR brightness peaks (Figure 6, left). A far-IR interferometer would observe these systems where they are brightest and with the required sub-arcsecond resolution (Figure 5 right, and 6), bringing hundreds of disks into structure-viewing range (Figure 7).[3]

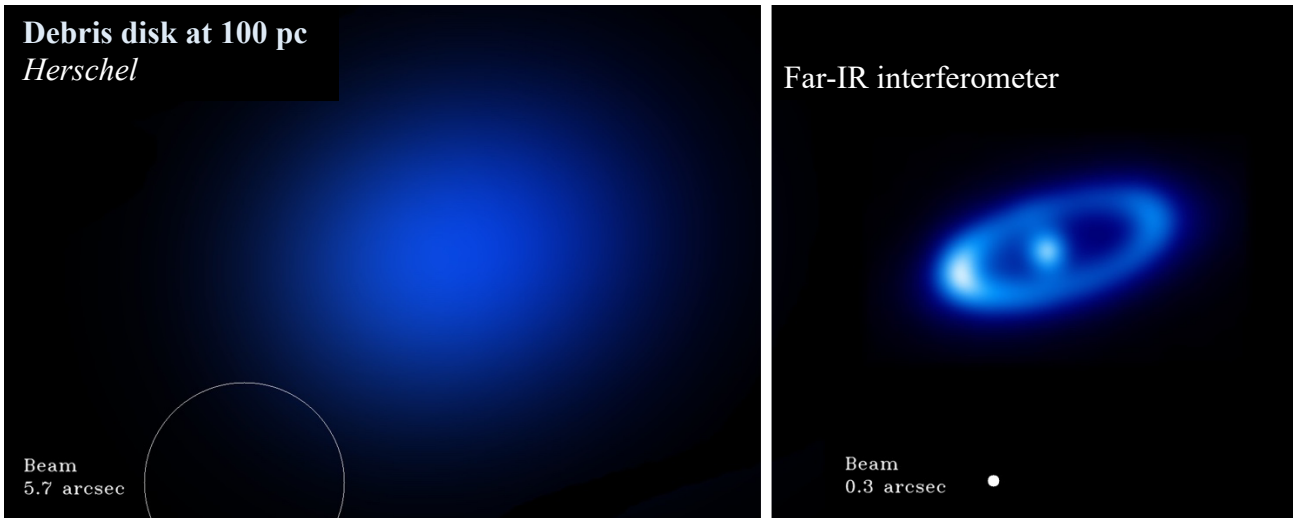

Figure 5. Most debris disks are located at distances too great for *Herschel* [10] to have resolved, and thus their structure is unknown. ALMA can resolve these disks but most of them are too faint to be seen at ALMA's submillimeter and millimeter wavelengths. For the sake of illustration, we show two images of a debris disk at the hypothetical distance d=100 pc. On the left is a *Herschel* 70-μm image, and on the right is a simulated 36-m interferometer image at 100 μm. Both images are based on an actual *Herschel* observation of Fomalhaut, one of very few nearby debris disks (d=7.7 pc), scaled to the angular size it would have at d=100 pc. With a far-IR interferometer we will be able to measure the structure in a large sample of debris disks.

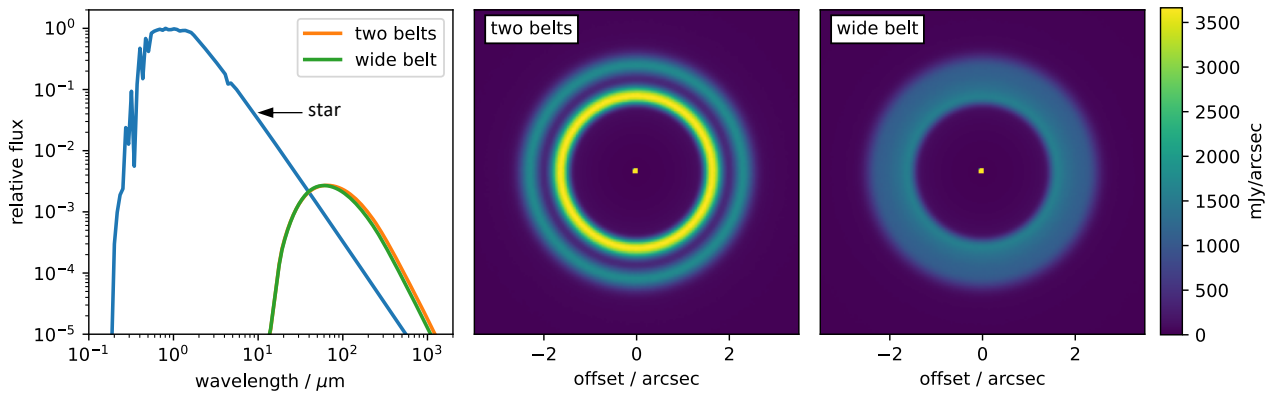

Figure 6. Two debris disk models from [3], both with the same amount of dust distributed between 30 and 50 au from the central star (right panels), and spectra for each of the two models (left). The spectral energy distribution gives a rough indication of the location of the dust belt but cannot be inverted to derive detailed structure (e.g., distinguishing between a single wide belt and two narrow belts). ALMA observes debris disks on the Rayleigh-Jeans tail of their spectral energy distribution, where the disks are orders of magnitude fainter than in the infrared. The longest wavelengths observable with JWST are insensitive to cold dust. A far-IR interferometer would operate in the spectral sweet spot where the disks are brightest and with resolution sufficient to see the disk structure.

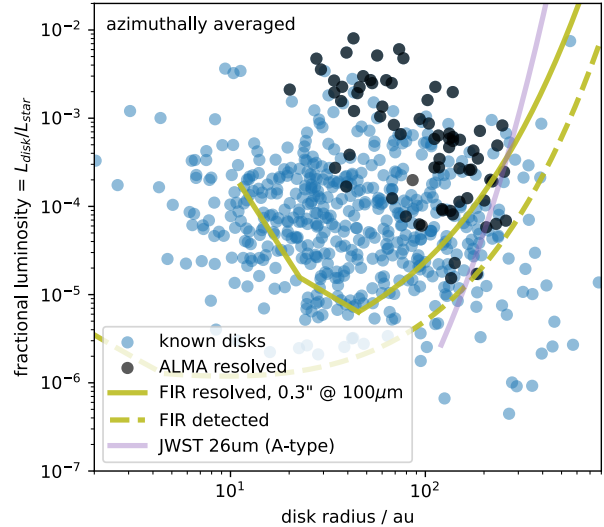

Figure 7. The vast majority of all known debris disks can be radially resolved at 0.3", the 100-μm angular resolution of an interferometer with maximum baseline length 36 m. ALMA resolves disks with the greatest fractional luminosity relative to their host stars. A far-IR interferometer would greatly expand the sample, enabling searches for trends in sub-populations (e.g., by stellar type and disk age). JWST resolves many of these disks, but its wavelength range traces small dust grains that are greatly influenced by non-gravitational forces and do not maintain the structure seen in their parent bodies. Ref. [3].

## 4. GALAXY EVOLUTION

We learn about galaxies through images at many wavelengths and by measuring their spectra. Images have informed our understanding of galaxy mergers and interactions, as well as morphological evolution and the prolonged existence of spiral structure. Much additional information is available in the spectral domain. A galaxy's overall spectral energy distribution from x-ray to radio wavelengths indicates the extent to which starlight is intercepted by interstellar dust and re-radiated in the far-infrared, the prevalence of star formation activity, and the presence of a dominant Active Galactic Nucleus (AGN). Spectral lines are excellent indicators of a galaxy's chemical composition (e.g., the relative abundance of "heavy elements" that can only be forged in stars, as opposed to the light primordial elements hydrogen and helium), physical conditions (e.g., temperature and density of the interstellar medium or the dominance of an AGN at the galaxy's center), and stellar population.

The spectra of galaxies are redshifted in proportion to their distance due to the expansion of the universe. Spectral lines are the most precise indicators of a galaxy's redshift, revealing not only its distance but also the corresponding look-back time, which represents the galaxy's place in cosmic history. JWST was needed to extend the reach of the *Hubble Space Telescope* farther back in time because the starlight in distant galaxies is redshifted to wavelengths inaccessible to *Hubble* but observable with JWST. Likewise, a galaxy's thermal dust emission is redshifted to wavelengths beyond the reach of JWST, and several spectral lines that serve as excellent diagnostics of the presence of a central supermassive black hole are redshifted and detectable at far-IR wavelengths.

*Hubble*, JWST, and ALMA have contributed greatly to our understanding of galaxy evolution, but sizeable gaps remain in our understanding due to the lack of a comparably powerful far-IR observatory. We have an incomplete picture of the evolution of cosmic star-formation history and would like to understand the co-evolution of galaxies and their central supermassive black holes, a US Decadal Survey[11] priority. The *Herschel Space Observatory* operated in the far-IR with a 3.5-m telescope and provided important insights, but its angular resolution, about 10", was such that the light of many galaxies appeared in each resolution element. Deblending the emission using spectral lines as markers is possible up to a point, but only with higher spatial resolution can far-IR galaxies can be identified with X-ray and optical counterparts and spectral energy distributions studied. To complete our understanding of the universe's star formation history and the co-evolution of galaxies and their central black holes, we need a sensitive observatory that can measure the far-IR spectra of individual distant galaxies. Figures 8 and 9 show that a colder and therefore more sensitive telescope than *Herschel* or a smaller single-aperture telescope will not have this measurement capability; sub-arcsecond resolution is an imperative.

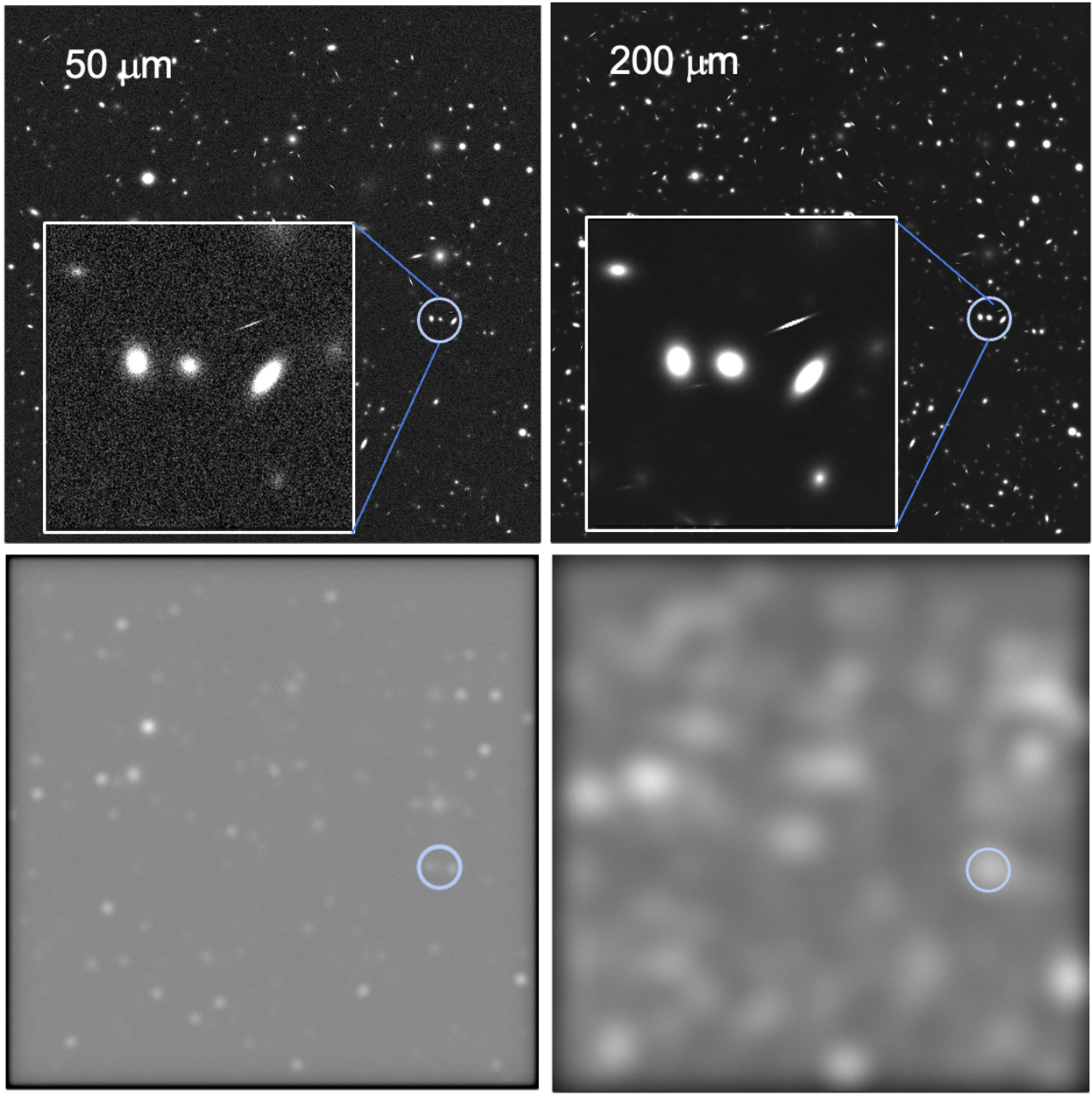

Figure 8. 50 and 200-μm wavelength slices from a simulated hyperspectral data cube (top row) convolved to the resolution of a 2-m telescope (bottom row) since such telescopes were and are under consideration. Each square image is 4.74 arcminutes on a side. A bright spot in the lower-right image (circle corresponding to a 2-m telescope's 200-μm beam) is chosen for closer inspection since observation with a single-aperture telescope might lead one to conclude it corresponds to "a galaxy." In fact, as shown in the inset images in the top panel, the circle circumscribes multiple bright galaxies. Simulation details will be presented in [4].

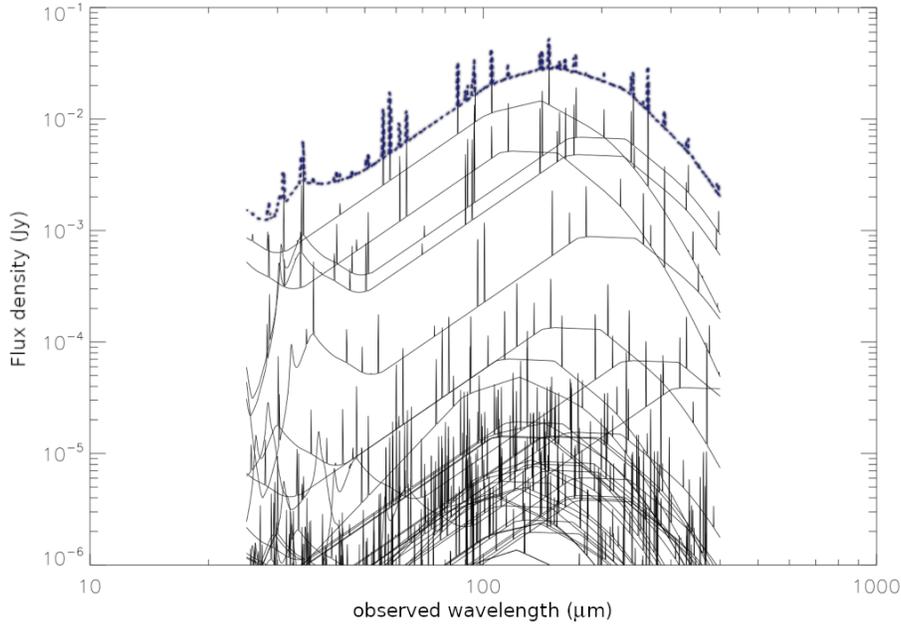

Figure 9. Spectra (solid lines) of the relatively bright galaxies that lie within the circles shown in Figure 8 and the sum of those spectra (dashed line), which represents the spectrum observable with a 2-m far-IR telescope. Even if the spectral lines from distinct galaxies are non-overlapping, their continuum emission and the mid-IR broadband feature from polycyclic aromatic hydrocarbons (PAHs) cannot be deblended. Aperture photometry was used to derive the individual galaxy spectra from the simulated data cube and a small region known to be free of simulated galaxies was used to derive the modeled foreground emission. The same foreground spectrum was subtracted from each of the individual galaxy spectra. Only an observatory with sub-arcsecond angular resolution can measure the galaxies as individual objects to study the evolution of galaxy properties throughout cosmic history.

## 5. FAR-IR MEASUREMENT REQUIREMENTS

We contend not only that far-IR observations, complementing those made with JWST and ALMA, are crucial to answering "How-did-we-get-here?" questions, but that access to sub-arcsecond angular scales in the far-IR is essential to expanding our knowledge horizon.

Over half of the energy emitted by the universe appears in the relatively unexplored far-IR spectral region, most of which is inaccessible from the ground due to the opacity of Earth's atmosphere, necessitating space-borne instrumentation. The ESA *Herschel Space Observatory*[10] (2009-2013) provided the best available views of the far-IR universe, serving to advance our understanding of the formation and evolution of planetary systems, stars, galaxies, and the universe as a whole. *Herschel* has shown what can be accomplished with a relatively large-aperture single-dish far-IR space telescope. Even with its success, *Herschel* observations are limited in angular resolution and sensitivity by its 3.5m diameter passively cooled primary mirror, which emits as an ~80 K graybody.

*Herschel* observations have also served to highlight the dramatically poorer angular resolution and sensitivity of far-IR facilities compared with that provided on either side of this spectral region[12]. The angular resolution provided by *Herschel* is comparable to that of Galileo's first telescopes (c. 1610). Many *Herschel* discoveries are waiting on enhanced spatial resolution follow-up observations to address questions like those raised in the preceding sections. JWST does not see out to long enough wavelengths, and ALMA cannot see through the atmosphere in the far-IR. A far-IR interferometer offers observations of astronomical objects in the spectral range where most of their light is emitted, access to critical diagnostic spectral lines, and measurements of water in its frozen and gaseous states,[2] all at angular resolution scales sufficient to resolve circumstellar disks[2,3] and penetrate extragalactic source confusion.[4] Attaining such angular resolution is simply not practical through any observational technique other than interferometry; to match JWST's resolution at 10x longer wavelengths with a single aperture would require a 10x larger (in diameter) telescope. Figure 10 illustrates the angular resolution capabilities of *Herschel*, the *Stratospheric Observatory for Infrared Astronomy* (SOFIA), the *Spitzer Space Telescope*, JWST, and ALMA, along with that which could be provided by a space-based far-IR interferometer mission employing spatial baselines of order tens of meters.

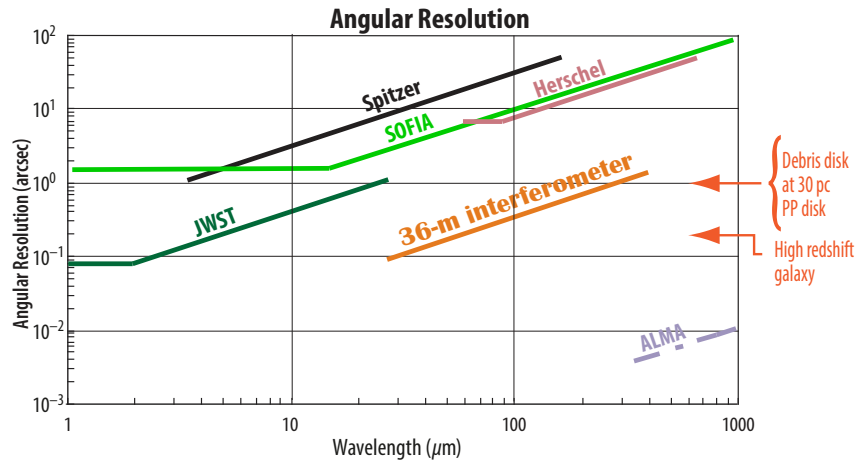

Figure 10. The angular resolutions of various past (*Spitzer*, SOFIA, *Herschel*) and present (JWST, ALMA) observatories are shown along with that of the SPIRIT[1] far-infrared interferometer (orange). For reference, arrows on the right-hand side indicate the angular resolution scale required to observe distant high-redshift galaxies and resolve nearby debris disks and protoplanetary disks.

While *Herschel's* primary mirror was passively cooled, the next-generation far-IR space telescope will employ actively cooled optics (~4-10 K) and state-of-the-art detectors such that observations are limited in sensitivity only by the astronomical background (predominantly zodiacal and Galactic cirrus dust emission at far-IR wavelengths). These technological advances will allow for a significant increase in the depth/speed of coverage within a given period of time, potentially allowing for probing of the distant cosmos, just as *Herschel* sensitively investigated the local universe. When cryo-cooled, even meter-class telescopes will provide exquisite sensitivity.

While enhanced angular resolution in astronomical imaging is critical to understanding the scene dynamics of regions under study, and, of course, it is the images that generate much widespread public interest in astronomy, spectral resolution is also key to understanding the astrophysics of a scene. A far-IR interferometer combines both angular and spectral resolution to provide the best of both worlds, with a modest compromise in the degree of complexity required to convert raw data into processed hyperspectral images. The significant improvements in sensitivity allowing for increase in mapping speeds can also help to compensate for the additional observational time/complexity required for interferometric observations. Figure 11 illustrates the expected spectral line sensitivity of a far-IR interferometer, given assumed detector NEPs of order $10^{-19}$ W Hz$^{-1/2}$ and cryogenic primary optics, *i.e.*, the astronomical photon noise-limited case, in the context of other far-IR and related facilities. Again, sensitivities of the order of JWST are provided at ~10x the wavelength.

With a Fourier transform spectrometer/interferometer, observed spectral resolution can be tuned to the target at hand, where observation time can be spent either on obtaining higher spectral resolution or on averaging for enhanced signal-to-noise ratio. The nature of a SPIRIT-like[1] "double Fourier" (spatio-spectral) interferometer is such that an additional dimension of spatial sampling is added to this parameter space. The spatial and spectral resolution profile is user-configurable to the observation target. The model far-IR interferometer under consideration has its wavelength coverage divided into 4 spectral bands, each with spectral resolution capabilities dictated by the maximum optical path difference available to the Fourier scanning element within the instrument, with increasing optical folds provided for the longer wavelength channels in order to better balance the overall spectral resolving power. The expected spectral resolving power is illustrated in Figure 12, along with those of the comparison observatories. Yet again, spectral resolving power similar to that of JWST is available at ~10x the wavelength.

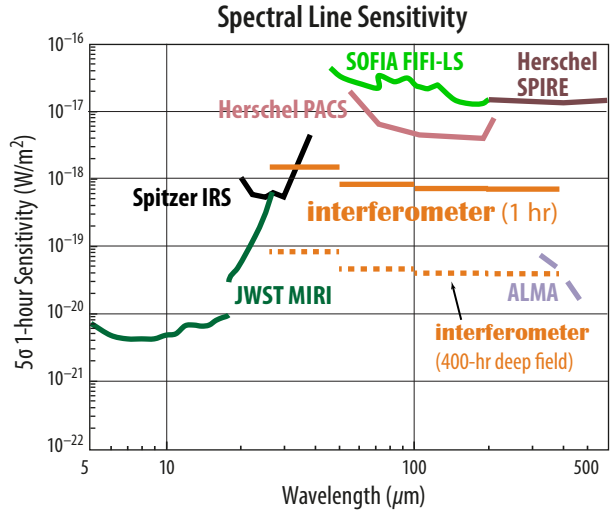

Figure 11. The 5σ, 1-hour spectral line sensitivity associated with the observatories presented in Fig. 10. The solid orange line indicates the expected sensitivity for a 1-hour observation with a far-IR interferometer, and the dashed line indicates that expected for a 400-hour deep field survey.

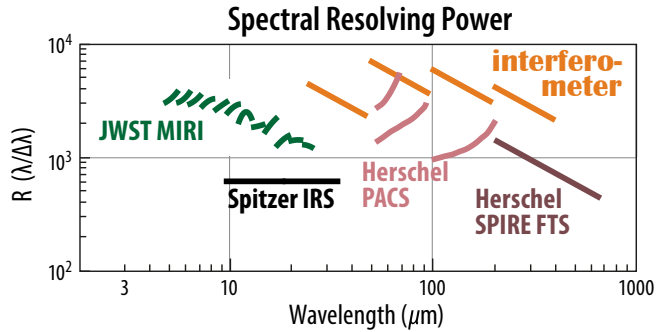

Figure 12. Maximum achievable spectral resolving power for the example observatories included in Figs. 10 and 11. The higher resolving power available through the far-IR interferometer under consideration is determined by the optical delay scan range in the beam-combining instrument.

For a given observational target, the dual spatial/spectral interferometry of a SPIRIT-like interferometer allows for operational optimization between dense vs. sparse baseline coverage (affecting image quality), longer or shorter integration time (affecting sensitivity), and longer or shorter delay scans (affecting spectral resolution and field of view).

## 6. CONCLUSION

In 2023, with no operating far-IR observatory and much to be learned through further investigations in this wavelength range, it is timely to consider the measurement capabilities most needed to achieve great advances in our understanding of the universe. Although both *Herschel* and *Spitzer* operated successful far-IR missions with high spectral resolution, they were limited in spatial resolution due to the fundamental diffraction limit of their 3.5-m and 0.85-m primary mirrors, respectively. Reaching sub-arcsecond resolution in the far-infrared is an observational imperative for reasons presented in this paper, as well as many others we have neglected. ALMA and JWST probe the sub-arcsecond regime at longer and shorter wavelengths, and a far-IR interferometer will do the same. Together these facilities will take us a long way down the path toward answering the question How did we get here?

*Acknowledgments*

A wonderful team gathered to develop the science case for the Space Interferometer for Cosmic Evolution (SPICE) NASA Probe mission concept. All SPICE team members contributed in one way or another to the ideas presented in this paper. In addition to the SPICE science team members who co-authored this paper, we thank Susanne Aalto (Chalmers University), Katey Alatalo (STScI), Jenny Bergner (UC Berkeley), Colm Bracken (Maynooth University), Kirstin Doney (Lockheed Martin), Steve Eales (Cardiff University), Matt Griffin (Cardiff University), Grant Kennedy (Warwick University), Al Kogut (NASA GSFC), Luca Matra (Trinity College Dublin), Taro Matsuo (Nagoya University), Melissa McClure (Leiden University), Joan Najita (NOIRLab), Nicole Pawallek (University of Vienna), Dave Sanders (University of Hawaii), Nick Scoville (Caltech), Laura Sommovigo (Scuola Normale Superiore di Pisa), Jessica Sutter (UCSD), Leon Trapman (University of Wisconsin Madison), Carole Tucker (Cardiff University), Gerard van Belle (Lowell Observatory), Serena Viti (Leiden University), Grant Wilson (UMass Amherst), and Mark Wyatt (University of Cambridge) for their generous time commitments and precious thoughts.